\documentstyle[a4]{elsartwb}
\input epsf
\input psfig

\begin{document}

\begin{frontmatter}

% \noindent \hfill {\bf nucl-th/0103027}
% \hfill INP yyyy/PH
\vspace{0mm}

\title{{\bf $\rho \to \pi \pi$ decay in nuclear medium}\thanksref{grants}}
\thanks[grants]{Research supported by PRAXIS grants XXI\-/BCC\-/429\-/94,
        PRAXIS\-/P\-/FIS\-/12247\-/1998, by
        Fundac\~{a}o para a Ci\^{e}ncia e a Tecnologia, POCTI/2000/FIS/35304,
        and by
        the Polish State Committee for
        Scientific Research grant 2P03B09419.}

\thanks[emails]{\hspace{0mm} E-mail:
Wojciech.Broniowski@ifj.edu.pl,
 brigitte@teor.fis.uc.pt,
 Woj\-ciech\-.\-Flor\-ko\-w\-ski\-@\-ifj.\-edu.\-pl}

\author[INP]{Wojciech Broniowski},
\author[INP]{Wojciech Florkowski},
\author[Coimbra]{Brigitte Hiller}

\address[INP]{The H. Niewodnicza\'nski Institute of Nuclear Physics,
         PL-31342 Cracow, Poland}

\address[Coimbra]{Centro de {F\'{\i}sica T\'eorica},
University of Coimbra, P-3004 516 Coimbra, Portugal}

\begin{abstract}
We calculate the medium modifications of the $\rho \pi \pi$ vertex
in a relativistic hadronic
framework incorporating nucleons and $\Delta(1232)$ isobars, and
find a substantial increase of the $\rho \pi \pi$
coupling, dominated by processes where the $\Delta$ is excited.
The coupling depends significantly on the virtuality of the $\rho$,
which is related to analytic properties of the vertex function.
We analyze the general case of a non-zero three-momentum
of the $\rho$ with respect to the nuclear
medium, and evaluate the resulting widths and spectral strength
in the transverse and
longitudinal channels for the $\rho \to \pi \pi$ decay.
These widths are used to obtain the dilepton
yields from $\rho$ decays in relativistic heavy-ion collisions.
\end{abstract}

\begin{keyword}
Mesons in nuclear medium, dilepton production in
relativistic heavy-ion collisions
\end{keyword}

\end{frontmatter}

\vspace{-7mm} PACS: 25.75.Dw, 21.65.+f, 14.40.-n

\section{Introduction\label{intro}}

In recent years a lot of efforts have been undertaken in order to understand
the properties of hadrons in nuclear medium. This challenging theoretical
issue gains a lot of importance at the beginning of the operation of RHIC,
where a proper and accurate inclusion of hadrons is necessary to describe
the evolution of the ``hadronic soup'' formed in the collision. It is
commonly accepted that hadrons are significantly modified by nuclear matter.
The arguments range from simple scaling of masses \cite{brscale,celenza},
through numerous hadronic model calculations \cite
{serot,chin,hatsuda,jean,herrmann1,herrmann2,herrmann3,pirner,%
Urban0,urban,rapp,Post,Peters,LeupoldRev,Gao}%
, QCD sum-rule techniques \cite{hatlee,leupold,lee2000}, approaches
motivated by the chiral symmetry \cite
{HatsudaKuni,Vogl,klingl,Cabrera1,Cabrera2,Chanfray,BH1,BH2}, to
model-independent predictions based on low-density expansion and dispersion
relations \cite{eletsky,FrimanActa,Lutz99,Friman2000}. In these calculations
masses of hadrons, or widths, or both, are significantly changed by the
presence of the medium (for a recent review see \cite{hadrons,tsuk,torino}).
In addition, medium may induce meson mixing absent in the vacuum \cite
{hindu1,BrFlmix,hindu2,hindu3}. An indirect indication for the modification
of vector mesons has been provided by the dilepton production measurements
in relativistic heavy-ion collisions (CERES \cite{ceres}, HELIOS \cite
{helios}). The dilepton productions has been theoretically studied with
medium-modified vector mesons \cite
{Dom,LiKoBrown1,LiKoBrown2,BratKo,Li2000,RappRev}, which helps to explain
the low-mass enhancement of the dilepton yields.

There is a lot of studies of the meson two-point functions in the
literature, but only a few devoted to meson three-point functions. In their
study of the $\rho $-meson in-medium spectral function Herrmann, Friman, and
N\"{o}renberg \cite{herrmann1} have computed the $\rho \pi \pi $ vertex for
the $\rho $ at rest in nuclear matter. They have applied a hadronic model
with the $\Delta $ isobar and with non-relativistic couplings. Temperature
effects on the $\rho \pi \pi $ interaction have been considered by Song and
Koch \cite{Song2}. Urban, Buballa, Rapp and Wambach \cite{Urban0,urban} have
extended the calculation of Ref. \cite{herrmann1} to non-zero
three-momenta of the $\rho $ meson and non-zero temperature. The authors of
the present paper have analyzed the $\omega \rightarrow \pi \pi $ decay in
nuclear medium \cite{BFH1,BFH2,BFH3}. Krippa has theoretically studied the
effects of density on chiral mixing of meson three-point functions \cite
{Krippa}. In Ref. \cite{RuHi} the $\rho \pi \pi $ coupling has been studied
in quark matter. Otherwise, the topic of medium effects on hadronic
couplings is very much {\em terra incognita}. Certainly, if the two-point
functions are significantly altered by the medium, one expects that the
three-point functions should also change. The issue is important for
modeling the hadronic evolution in relativistic heavy-ion collisions, since
the change of the hadron couplings results in altered transition rates
between hadrons.

Out of many meson couplings, the $\rho \pi \pi $ vertex is especially
important, since the $\rho $ plays an essential role in hadron dynamics. The
vacuum value of the coupling constant is large, $g_{\rho \pi \pi }\simeq 6$.
Since the couplings of $\rho $ and pions to nucleons and $\Delta $ isobars
are also large, we expect significant medium 
modifications of $g_{\rho \pi \pi }$. 
As our calculation shows, this is indeed the case, with the
coupling significantly altered already at the nuclear saturation density.
The effect is mainly due to the interactions with the $\Delta $. Our method
is similar to Refs. \cite{herrmann1,Urban0}, with the following differences:
We use a {\em fully relativistic} framework, with relativistic interactions.
The {\em leading-density} approximation is used, which makes the calculation
very simple, as no integration over the nucleon momenta is necessary. We
analyze analyticity in the virtual $\rho $ mass, which is non-trivial in our
problem due to low-mass thresholds. These thresholds largely influence
the results. In addition, keeping non-zero three-momentum of the $\rho $, as
in Ref. \cite{Urban0}, allows us to look separately on the longitudinal and
transverse polarizations. In our comparison to the CERES dilepton data we
use the fire-cylinder expansion model of Refs. \cite{Rappevol,RappShur} and
take into account the experimental cuts, which are very important for the
detailed numerical analysis. 

The outline and the main results of the paper are as follows: In Sect. \ref
{frame} we introduce our framework,{\em \ i.e.} the relativistic model of
mesons interacting with nucleons and $\Delta $ isobars, the latter treated
as Rarita-Schwinger fields. We introduce the necessary hadronic vertices and
set the coupling constants. As already mentioned, we work at zero
temperature and to the leading-order in baryon density. In Sect. \ref{q0} we
present our results for the $\rho \pi \pi $ vertex for the $\rho $ meson at
rest with respect to the medium, and point out large medium effects even at
the nuclear saturation density. We investigate analyticity of the coupling
in the virtual $\rho $ meson mass and point out its significance for the
quantitative results. We look at effects of the energy-dependent width of the $%
\Delta $ and find they are not very large. In Sect. \ref{moving} we analyze
the general case of finite three-momentum of the $\rho $ with respect to the
medium, and calculate the widths and spectral functions in the transverse
and longitudinal channels. In Sect. \ref{dilep} we apply the obtained
spectral functions to compute the dilepton production rate from $\rho $
decays for the CERES $Pb+Au$ experiment. The calculation carefully includes
the CERES experimental kinematic cuts, which is crucial for the results. We
also include the effects of the expansion of the fire cylinder, which enter
at the level of a few percent. It is found that spreading of the width helps
to explain the low-mass enhancement of the dilepton spectra, but the overall
normalization fails significantly short of the experimental data. Section 
\ref{summa} presents our conclusions and a discussion of several additional
points.

\section{The framework\label{frame}}

Throughout this paper we choose conventionally $q$ as the incoming momentum
of the $\rho $ carrying isospin index $b$ and polarization vector $%
\varepsilon^{\mu }$, $p$ as the outgoing momentum of the pion carrying isospin
index $a$, and $q-p $ as the outgoing momentum of the pion carrying isospin
index $c$. With this convention, the vacuum value of the $\rho \pi \pi $
vertex (Feynman rule) is 
\begin{equation}
-iV_{\rho _{\mu }^{b}\pi ^{a}\pi ^{c}}=g_{\rho }\epsilon ^{acb}(2p^{\mu
}-q^{\mu }).  \label{rppvac}
\end{equation}
The medium modifications of the coupling, calculated below, will be compared
to Eq. (\ref{rppvac}).

Our calculation of the in-medium $\rho \pi \pi $ vertex is made in the
framework of a fully 
relativistic hadronic theory, where mesons interact with the
nucleons and $\Delta $ isobars. Other in-medium calculations indicate that
the leading-density approximation is sufficient up to densities of the order
of the nuclear saturation density.\footnote{An extension beyond the
leading-density approximation would require accounting for the Fermi-sea
motion, which technically leads to keeping the integration over ${\bf k}$ in
our expressions, and, more importantly, the inclusion of 
correlations. In many similar calculations the Fermi-motion effects have
been found to influence the results weakly at moderate densities. The 
inclusion of nucleon correlations is a difficult problem, 
extending far beyond the present
work.} Also, in Ref. \cite{urban} the finite-temperature effects in the $%
\rho \pi \pi $ coupling have been found to be moderate for temperatures up
to about $\sim 150$MeV, typical for relativistic heavy-ion collisions (cf.
Fig. 7(a,b) of \cite{urban}). This justifies the use of zero temperature
to obtain the vertices of Fig. 1, at least as a
first approximation. 

To the leading-density order only the diagrams shown in Fig. 1 contribute to
the $\rho \rightarrow \pi \pi $ process. The solid lines denote $i$ times
the in-medium nucleon propagator, which can be decomposed in the usual way
into the {\em free} and {\em density} parts \cite{chin}: 
\begin{eqnarray}
iG(k) &\equiv &iG_{F}(k)+iG_{D}(k)=  \label{Nprop} \\
&=&i(\FMslash{k}+m_{N})\left[ \frac{1}{k^{2}-m_{N}^{2}+i\varepsilon }+\frac{%
i\pi }{E_{k}}\delta (k_{0}-E_{k})\theta (k_{F}-|{\bf k}|)\right] ,  \nonumber
\end{eqnarray}
where $k$ denotes the nucleon four-momentum, $m_{N}$ is the nucleon mass, $%
E_{k}=\sqrt{m_{N}^{2}+{\bf k}^{2}}$, and $k_{F}$ is the Fermi momentum of
nuclear matter. The double line in the diagrams of Fig. 1 denotes $i$ times
the relativistic $\Delta $ propagator, 
\begin{eqnarray}
iG_{\Delta }^{\alpha \beta }(k) &=&i{\frac{\FMslash{k}+m_{\Delta }}{%
k^{2}-\left( m_{\Delta }-\frac{i}{2}\,\Gamma _{\Delta }\right) ^{2}}}\left(
-g^{\alpha \beta }+{\frac{1}{3}}\gamma ^{\alpha }\gamma ^{\beta }+{\frac{%
2k^{\alpha }k^{\beta }}{3m_{\Delta }^{2}}}+{\frac{\gamma ^{\alpha }k^{\beta
}-\gamma ^{\beta }k^{\alpha }}{3m_{\Delta }}}\right) .  \nonumber \\
&&  \label{Dprop}
\end{eqnarray}
This formula corresponds to the usual Rarita-Schwinger definition \cite
{rarita,Ben}, with the parameter choice $A=-1$. We have modified the
denominator in Eq. (\ref{Dprop}) in order to account for the finite width of
the $\Delta $ resonance.
% fixing $\Gamma _{\Delta }$ at the vacuum value of $%120$MeV.

Since we are interested in density effects, one of the nucleon lines in each
of the diagrams of Fig. 1 must involve the nucleon density propagator, $%
G_{D} $. For kinematic reasons, diagrams with more than one $G_{D}$ vanish.
The wavy line in Fig. 1 denotes the $\rho $ meson, and the dashed lines
correspond to pions. All external particles are on mass shell.
\begin{figure}[t]
\centerline{\psfig
{figure=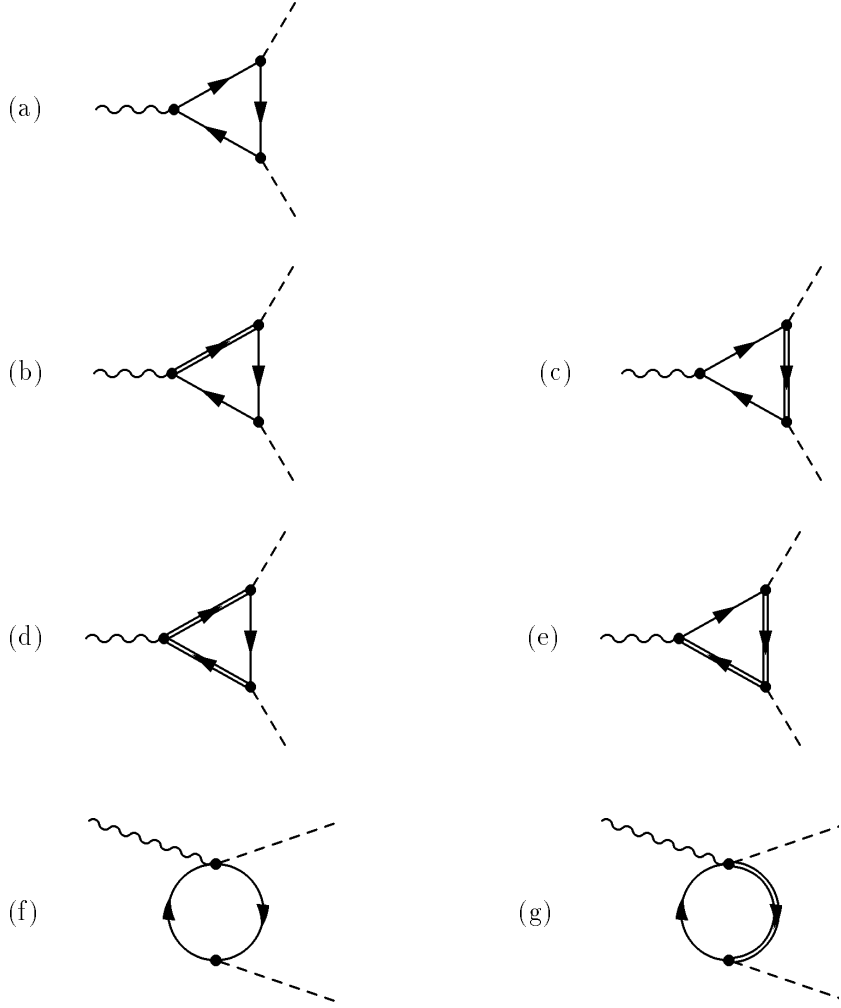,height=14cm,bbllx=82bp,bblly=219bp,bburx=456bp,bbury=677bp,clip=}}
\label{diag}
\caption{Diagrams included in our calculation (crossed diagrams not
displayed). Wavy lines denote the $\rho$ meson, dashed lines the
pions, solid lines the in-medium nucleon, and double lines the $\Delta$.}
\end{figure}

The meson-nucleon vertices (Feynman rules) needed for our calculation have
the standard form 
\begin{eqnarray}
-iV_{\pi ^{a}NN} &=&\frac{g_{A}}{2F_{\pi }}\FMslash{p}\gamma _{5}\tau ^{a},
\label{VpiNN} \\
-iV_{\rho _{\mu }^{b}NN} &=&ig_{\rho }(\gamma ^{\mu }+\frac{i\kappa _{\rho }%
}{2m_{N}}\sigma ^{\mu \nu }q_{\nu })\frac{\tau ^{b}}{2},  \label{VrhoNN} \\
-iV_{\rho _{\mu }^{b}\pi ^{a}NN} &=&i\frac{g_{\rho }g_{A}}{2F_{\pi }}\gamma
^{\mu }\gamma _{5}\varepsilon ^{abc}\tau _{c}  \label{VrhopiNN}
\end{eqnarray}
where $p$ is the outgoing four-momentum of the pion, $q$ is the incoming
momentum of the $\rho $ meson, and $a$ and $b$ are the isospin indices of
the pion and the $\rho $, respectively. We have chosen the pseudovector
pion-nucleon coupling. The vertex (\ref{VrhopiNN}) follows from the minimal
substitution in Eq. (\ref{VpiNN}).

Whereas the couplings of mesons to nucleons listed above are well
established and the corresponding parameters are well known (except for the
long-lasting controversy with $\kappa _{\rho }$), the relativistic couplings
involving the $\Delta $ resonance are a topic of an on-going research and
discussion \cite{Ben,Pascal,Hemmert,Haber}. Several structures are possible
for these vertices, and ambiguities related to the choice of the so-called
off-shell parameters have not been resolved. For our pragmatic goal of
estimating the size of the in-medium $\rho \pi \pi $ vertex, we will adopt
the following simple-minded and popular philosophy: all off-shell couplings
are set to zero.

The meson-nucleon-$\Delta $ vertices have the following form \cite{Ben} 
\begin{eqnarray}
-iV_{\pi ^{a}N\Delta _{\alpha }} &=&\frac{{f_{\pi N\Delta }}}{{m_{\pi }}}%
\theta ^{\alpha \nu }(Z)p_{\nu }{T}^{a},  \label{VpiND} \\
-iV_{\rho _{\mu }^{b}N\Delta _{\alpha }} &=&ig_{1}(\theta ^{\mu \alpha }(Y)%
\FMslash{q}\gamma _{5}-\theta ^{\mu \nu }(Y)q_{\nu }\gamma ^{\alpha }\gamma
_{5})T^{b}  \nonumber \\
&&+ig_{2}(\theta ^{\mu \alpha }(X)\gamma _{5}k\cdot q-\theta ^{\mu \nu
}(X)q_{\nu }\gamma _{5}k^{\alpha })T^{b},  \label{VrhoND} \\
-iV_{\rho _{\mu }^{b}\pi ^{a}N\Delta _{\alpha }} &=&i\frac{g_{\rho }{f_{\pi
N\Delta }}}{{m_{\pi }}}\theta ^{\alpha \mu }(Z)\varepsilon ^{abc}{T}_{c},
\label{VrhopiND}
\end{eqnarray}
where $k$ is the nucleon four momentum, and 
\begin{equation}
\theta ^{\sigma \lambda }(W)=g^{\sigma \lambda }-(W+\frac{1}{2})\gamma
^{\sigma }\gamma ^{\lambda },\qquad W=Z,Y,X.  \label{ThetaRS}
\end{equation}
The combination $-(W+\frac{1}{2})$ is called the off-shell parameter.
According to the prescription stated above, we set $Z=Y=X=-\frac{1}{2}$,
such that all off-shell parameters vanish, and 
we simply have $\theta ^{\sigma \lambda
}(W)=g^{\sigma \lambda }$. Furthermore, we take arbitrarily $g_{2}=0$. The
matrices $T^{a}$ in Eqs. (\ref{VpiND}-\ref{VrhopiND}) are the standard
isospin ${\frac{1}{2}}\rightarrow {\frac{3}{2}}$ transition matrices, given
in App. \ref{rar}. The vertex (\ref{VrhopiND}) follows from the minimum
substitution in Eq. (\ref{VpiND}).

For the $\pi \Delta \Delta $ coupling there are, according to Ref. 
\cite{Hemmert}, three possible structures: 
\begin{equation}
iV_{\pi ^{a}\Delta _{\alpha }\Delta _{\beta }}=\left( G_{1}g^{\alpha \beta }%
\FMslash{p}\gamma _{5}+G_{2}(\gamma ^{\alpha }p^{\beta }+p^{\alpha }\gamma
^{\beta })\gamma _{5}+G_{3}\gamma ^{\alpha }\FMslash{p}\gamma _{5}\gamma
^{\beta }\right) T_{\Delta }^{a}.  \label{VpiDD}
\end{equation}
We drop the off-shell couplings by setting $G_{2}=G_{3}=0$. For the $\rho
\Delta \Delta $ vertex we use the minimal vector current coupling \cite{Ben}
and the universality prescription, which gives

\begin{equation}
iV_{\rho _{\mu }^{b}\Delta _{\alpha }\Delta _{\beta }}=ig_{\rho }(-\gamma
^{\mu }g^{\alpha \beta }+g^{\alpha \mu }\gamma ^{\beta }+g^{\beta \mu
}\gamma ^{\alpha }+\gamma ^{\alpha }\gamma ^{\mu }\gamma ^{\beta })T_{\Delta
}^{b}.  \label{VrhoDD}
\end{equation}
The constants $g_{1}$ and $G_{1}$ are adjusted in such a way, that in the
nonrelativistic limit we recover the couplings $\sqrt{2}({f}^{\ast }{/m}%
_{\pi }){\varepsilon }_{ijk}p^{i}{S}_{\Delta }^{j}{T}_{\Delta }^{a}$ and ${%
(f_{\Delta }/m}_{\pi })p_{j}{S}_{\Delta }^{j}{T}_{\Delta }^{a}$,
respectively \cite{EW,Gomez}, where $k$ is the spin index of the $\rho$. The
comparison, with the explicit form of the Rarita-Schwinger spinors and the
matrices $T_{\Delta }$ and $S_{\Delta }$ (see App. \ref{isoalg}) gives $%
G_{1}=$ $\frac{3}{2}{f_{\Delta }/m}_{\pi }$ and $g_{1}=$ $\sqrt{2}{f}^{\ast }%
{/m}_{\pi }$. Our choice of the physical parameters is as follows \cite
{Gomez}: 
\begin{eqnarray}
g_{A} &=&1.26,\qquad F_{\pi }=93{\rm MeV},\qquad m_{\pi }=139.6{\rm MeV}, 
\nonumber \\
g_{\rho } &=&5.26,\qquad \kappa _{\rho }=6,  \label{param} \\
f_{\pi N\Delta } &=&2.12,\qquad f^{\ast }=2.12,\qquad f_{\Delta }=0.802. 
\nonumber
\end{eqnarray}

In addition to the mentioned ambiguities in the choice of the form
of the $\Delta$ couplings \cite{Ben,Pascal,Hemmert,Haber}, 
there are uncertainties in the values of the coupling constants. For some
constants one typically uses the quark model predictions, or the large-$%
N_{c} $ arguments which relate the $\Delta $ couplings to the nucleon
couplings \cite{Dashen1,Dashen2,WBNC}. Any adopted scheme should fit the
values of the coupling constants to the available data for various
processes. However, for our purpose of estimating the size of the medium
effect on the $\rho \pi \pi $ vertex these ambiguities are not essential. In
addition, the result, as we shall see shortly, is dominated by diagram (g)
of Fig. 1, containing 
only the $\pi N\Delta $ and $\pi \rho N\Delta $ couplings which are
well established. Our results are not very sensitive to the choice of other
couplings. Also, for simplicity of the approach and from the 
lack of knowledge we do not include any form-factors in the vertices.

We work in the rest frame of the nuclear matter, however effort is made to
write all expressions covariantly, which turns out to be very convenient.
Our calculation is made in the following way. First, we evaluate the
diagrams of Fig. 1. The result for the full $\rho \rightarrow \pi \pi $
amplitude has the generic form ${\cal M}_{acb}=\varepsilon_{\mu
}A_{acb}^{\mu }$, where the vertex function is 
\begin{eqnarray}
A_{acb}^{\mu } &=&\epsilon ^{acb}(A_{{\rm vac}}^{\mu }+A_{{\rm med}}^{\mu }),
\label{Amu} \\
A_{{\rm vac}}^{\mu } &=&g_{\rho }(2p^{\mu }-q^{\mu }),  \nonumber \\
A_{{\rm med}}^{\mu } &=&\int \frac{d^{3}k\,}{(2\pi )^{3}} \frac{m_N}{E_k}%
(Ap^{\mu }+Bq^{\mu }+Ck^{\mu })\theta (k_{F}-|{\bf k}|),  \nonumber
\end{eqnarray}
and $A$, $B$, and $C$ are scalar functions depending on scalar products of
the four-vectors $q$, $p$, and $k$, with $k^{0}=E_{k}$. The occupation
function is made explicitly Lorentz-invariant when we write $|{\bf k}|=\sqrt{%
(k \cdot u)^2 -m_N^2}$, where $u$ is the four-velocity of the medium. The
term with $k^{\mu }$, upon the evaluation of the integral, can be in general
proportional to the three Lorentz vectors present in the problem, namely 
\begin{equation}
\int \frac{d^{3}k\,}{(2\pi )^{3}}\frac{m_N}{E_k}Ck^{\mu } \theta (k_{F}-|%
{\bf k}|)=C_{p}p^{\mu }+C_{q}q^{\mu}+C_{u}u^{\mu },  \label{decomp}
\end{equation}
where $C_{q}$, $C_{p}$,
and $C_{u}$ are scalar functions of $p^{2}$, $q^{2}$, $p\cdot q$, $p\cdot u$%
, $q\cdot u$, and $k_F$. Contracting Eq. (\ref{decomp}) with $p_{\mu }$, $%
q_{\mu }$, and $u_{\mu }$ we obtain a set of linear algebraic equations for $%
C_{q}$, $C_{p}$, and $C_{u}$, which can be solved. However, at the{\em \
leading-density }approximation the problem becomes even simpler. We can work
in the {\em rest frame of the medium}, where $u^{\mu }=(1,0,0,0)$. It is
obvious that the leading-density approximation is equivalent to setting the
three-vector ${\bf k}$ to zero in the functions $A$, $B$, $C$, and $E_k$
appearing in the integrands of Eqs. (\ref{Amu},\ref{decomp}). Then $\int 
\frac{d^{3}k\,}{(2\pi )^{3}}\theta (k_{f}-|{\bf k}|)=\frac{1}{4}\rho _{B}$,
while higher-order terms in ${\bf k} $ generate terms with higher exponents
of the baryon density $\rho _{B}$. Now, with ${\bf k}=0$, $k^{0}=m_{N}$, the
contraction of Eq. (\ref{decomp}) with $q_{\mu }$, $p_{\mu }$, and $u_{\mu }$
gives the set of equations 
\begin{eqnarray}
\frac{1}{4}\rho _{B}m_{N}p^{0} \overline{C} &=&C_{p}p^{2}+C_{q}q\cdot
p+C_{u}p^{0},  \nonumber \\
\frac{1}{4}\rho _{B}m_{N}q^{0} \overline{C} &=&C_{p}p\cdot
q+C_{q}q^{2}+C_{u}q^{0},  \label{set} \\
\frac{1}{4}\rho _{B}m_{N} \overline{C} &=&C_{p}p\cdot u+C_{q}q\cdot u+C_{u},
\nonumber
\end{eqnarray}
where $\overline{C}$ is $C$ with ${\bf k}=0$. Since in the general case
vectors $p$, $q$, and $u$ are linearly-independent, the solution of Eqs. (%
\ref{set}) is $C_{p}=C_{q}=0$, $C_{u}=\frac{1}{4}\rho _{B}m_{N}\overline{C}$%
. Thus, only the term proportional to $u^{\mu }$ in Eq. (\ref{decomp}) is
present in the leading-density approximation. In other words, we have 
\begin{equation}
A_{{\rm med}}^{\mu }=\frac{1}{4}\rho _{B}(\bar{A}p^{\mu }+\bar{B}q^{\mu }+%
\bar{C}m_{N}u^{\mu }),  \label{Amumed}
\end{equation}
where the coefficients $\bar{A}$, $\bar{B}$, and $\bar{C}$ are obtained from 
$A$, $B$, and $C$ by simply setting ${\bf k}=0$. 

To summarize this part, we restate the necessary steps needed to obtain 
the leading-density amplitude: The traces in diagrams of Fig. 1 are evaluated,
leading to Eq. (\ref{Amu}). Then we replace $k^\mu$ by $m_N u^\mu$, set 
${\bf k}=0$, and arrive at Eq. (\ref{Amumed}). Certainly, this 
very simple method is 
general for any problem involving baryon loops with density-dependent 
nucleon propagators.   
In our calculation we have
used a standard Dirac algebra package \cite{HIP}. The isospin traces are
evaluated in App. \ref{isoalg}.

\section{Results for $\rho $ decaying at rest\label{q0}}

We begin the presentation of the results with the case where the $\rho $ is
at rest with respect to nuclear matter, ${\bf q}=0.$ In this kinematics we
find $\bar{B}=-\frac{1}{2}\bar{A}$, and $\bar{C}=0$. The fact that $\bar{C}=0
$ is reflecting the equality of the transversely and longitudinally
polarized $\rho $ meson propagators at ${\bf q}=0$, as will be shown in Sect.
\ref{moving}. The result $\bar{B}=-\frac{1}{2}\bar{A}$ is consistent with
the Ward-Takahashi identity $q_{\mu }A^{\mu }=D_{\pi }^{-1}(p)-D_{\pi
}^{-1}(p-q)$, where $D_{\pi }$ denotes the pion propagator dressed with
nucleon and nucleon-$\Delta $ bubbles \cite{herrmann1}. With ${\bf q}=0$ and
pions on the mass shell, we 
obviously have $D_{\pi }^{-1}(p)=D_{\pi }^{-1}(p-q)$, and,
consequently, $\bar{A}p\cdot q+\bar{B}q^{2}=0$, which immediately gives $%
\bar{B}=-\frac{1}{2}\bar{A}$. Therefore, for ${\bf q}=0$ the in-medium
vertex function is proportional to $2p^{\mu }-q^{\mu }$: 
\begin{equation}
A_{{\rm med}}^{\mu }({\bf q}=0)=\frac{1}{8}\rho _{B}\bar{A}({\bf q}%
=0)(2p^{\mu }-q^{\mu }).  \label{mivf}
\end{equation}
The full formula for $\bar{A}({\bf q}=0)$ is very lengthy, hence we present
below only the contribution from the bubble diagrams (f,g), which are
simple. In App. \ref{ampliexp} we 
list the contributions to $\bar{A}({\bf q}=0)$
from all diagrams in the special case of $m_{\pi }=0$.% 
\footnote{It should be stressed that there is no obvious expansion 
parameter in the problem. Expanding the amplitude 
in $m_\pi$ or $m_\Delta- m_N$ and keeping
the lowest terms does not lead to a good approximation.} 
It turns out that the 
$N-\Delta $ bubble diagram (g) is the dominant one. We have
\begin{eqnarray}
\bar{A}^{(g)}({\bf q}=0) &=&\frac{16}{9}g_{\rho }\left( \frac{{f_{\pi
N\Delta }}}{{m_{\pi }}}\right) ^{2}\times   \label{Af} \\
&&\frac{2(m_{\Delta }^{2}-m_{\pi }^{2}-\frac{1}{2}m_{N}M)(m_{N}+m_{%
\Delta }+\frac{1}{2}M)}{m_{\Delta }^{2}(m_{N}^{2}-(m_{\Delta }-\frac{i}{2}%
\Gamma _{\Delta })^{2}+m_{\pi }^{2}+m_{N}M)}+(M\rightarrow -M),  \nonumber
\end{eqnarray}
with $M$ denoting the mass of the $\rho $ meson. For comparison, the
contribution from the $N-N$ bubble, Fig. 1(f), is
\begin{equation}
\bar{A}^{(f)}({\bf q}=0)=g_{\rho }\left( \frac{{g}_{A}}{{2F}_{\pi }}\right)
^{2}\frac{32m_{N} m_{\pi }^{2}}{m_{\pi }^{4}-m_{N}^{2}M^{2}}.  \label{Ae}
\end{equation}
In the limit of large $m_{N}$, with $m_{\Delta }-m_{N}$ fixed, and $\Gamma
_{\Delta }=0$, expression (\ref{Af}) reduces to 
\begin{equation}
\bar{A}^{(g)}({\bf q}=0)\rightarrow \frac{16}{9}g_{\rho }\left( \frac{{%
f_{\pi N\Delta }}}{{m_{\pi }}}\right) ^{2}\frac{4(m_{\Delta }-m_{N})}{%
M^{2}/4-(m_{\Delta }-m_{N})^{2}},  \label{limit}
\end{equation}
which agrees with non-relativistic calculations.

\begin{figure}[tbp]
\vspace{0mm} \epsfxsize = 11 cm \centerline{\epsfbox{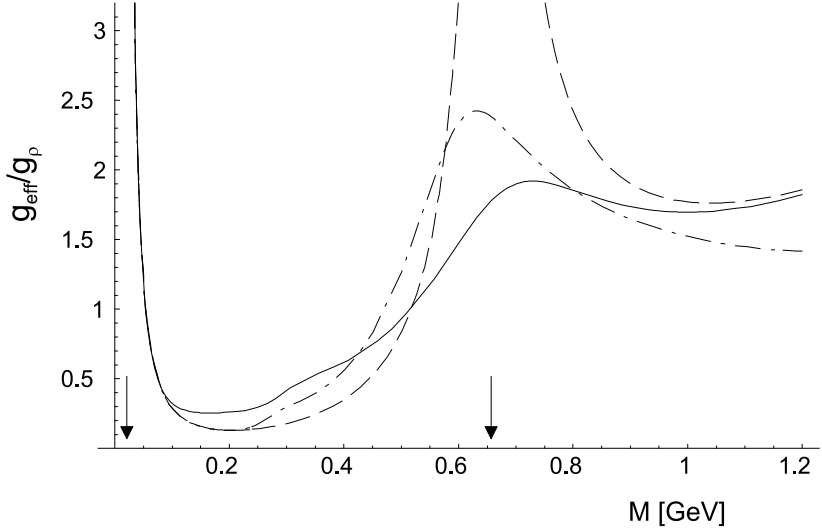}} \vspace{0mm} 
\label{figgeff}
\caption{The ratio of the effective coupling constant, Eq. (\protect\ref{geff}), at
the saturation density, to its vacuum value $g_\rho $,
plotted as a function of the virtual mass of the $\rho $, $M$. The
solid and dashed lines correspond to the case with $\Gamma _\Delta =120$%
MeV and $\Gamma _{\Delta }=0$, respectively. The dot-dashed lines 
corresponds to the energy-dependent $\Gamma _\Delta $ of
Eq. (\protect\ref{gedep}).  Arrows indicate positions of
the singularities of Eq. (\protect\ref{sin1}-\protect\ref{sin2}).}
\end{figure}

In the following we shall treat $M$ as the mass of a {\em virtual} $\rho $
meson. Virtual $\rho $ mesons are needed for the analysis of the dilepton
production in Sect. \ref{dilep}. Analyticity of the vertex function is
nontrivial in the variable $M$. We can see from the denominators of Eqs. (%
\ref{Af},\ref{Ae}), that for $\Gamma _{\Delta }=0$ singularities occur at 
\begin{eqnarray}
M^{2} &=&\left( \frac{m_{\Delta }^{2}-m_{N}^{2}-m_{\pi }^{2}}{m_{N}}\right)
^{2}=(0.657{\rm GeV})^{2},  \label{sin1} \\
M^{2} &=&\left( \frac{m_{\pi }^{2}}{m_{N}}\right) ^{2}=(0.021{\rm GeV})^{2}.
\label{sin2}
\end{eqnarray}
Triangle diagrams of Fig. 1 also have the above singularities, and
additionally bring in high-lying 
particle-antiparticle production
singularities at $M^{2}=(2m_{N})^{2}$ and
at $M^{2}=\left( m_{\Delta }+m_{N}\right) ^{2}$, which are physically not
relevant. The above analytic structure is manifest in the numerical
calculations presented below. For non-zero $\Gamma _{\Delta }$ the pole at (%
\ref{sin1}) changes to a broader structure. Thus, analyticity is important
--- it immediately leads to large changes of the vertex function near the
poles.

\begin{figure}[tbp]
\vspace{0mm} \epsfxsize = 11 cm \centerline{\epsfbox{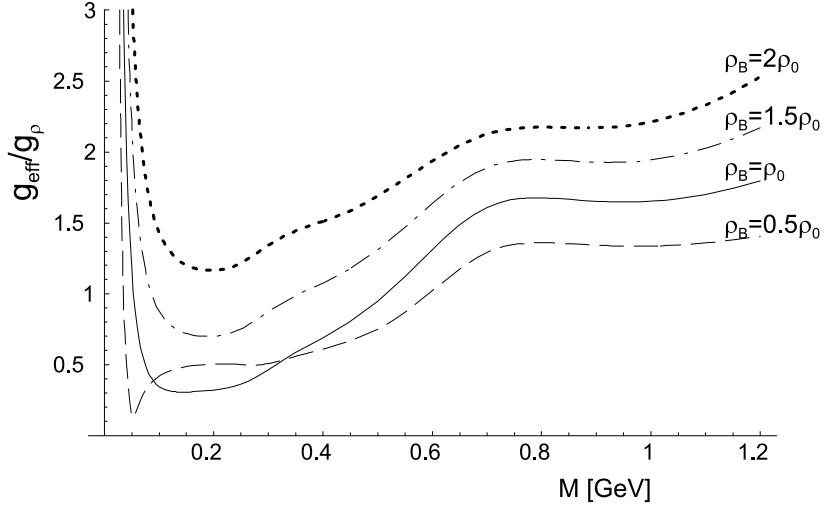}} \vspace{0mm}
\label{geff2}
\caption{Same as Fig. 2 for different values of the baryon density, 
$\rho _{B}$, with $\Gamma _{\Delta }$ depending on density 
according to parameterization (\protect\ref{rhopara}). }
\end{figure}

Since for the case of ${\bf q}=0$ the matter-induced vertex function (\ref
{mivf}) has the same Lorentz structure as in the vacuum,{\em \ i.e.}
proportional to $2p^{\mu }-q^{\mu }$, it is convenient for our quantitative
studies to introduce an effective $\rho \pi \pi $ coupling constant, defined
as 
\begin{equation}
g_{{\rm eff}}=\left| g_{\rho }+\frac{1}{8}\rho _{B}\bar{A}({\bf q}=0)\right|
.  \label{geff}
\end{equation}
The absolute value is taken, since with non-zero $\Gamma _{\Delta }$ the
quantity $\bar{A}({\bf q}=0)$ is complex. In Fig. 2 we plot the ratio $g_{%
{\rm eff}}/g_{\rho }$ as a function of the virtual $\rho $ mass, $M$, at the
saturation density, $\rho _{B}=\rho _{0}$. The dashed line corresponds to
the zero-width $\Delta $. We can clearly see the singularities of Eq. (\ref
{sin1},\ref{sin2}), whose positions are indicated by arrows. The solid line
in Fig. 2 shows the calculation with the vacuum value of the $\Delta $
width, $\Gamma _{\Delta }=120{\rm MeV}.$ In this case the pole at (\ref{sin1}%
) changes to a broad structure. The pole of Eq. (\ref{sin2}) remains, of
course, unchanged. The considerable difference between the solid and dashed
curves in the range of $M$ between $0.6$ and $1$GeV shows that the results
are sensitive to the assumed value for $\Gamma _{\Delta }.$ We note that at
low masses $M$, between $\sim 0.07$ and $\sim 0.55{\rm GeV}$, the effective
coupling $g_{{\rm eff}}$ is lower than the vacuum value, thus the medium
lowers the coupling, while above $M\sim 0.55{\rm GeV}$ the effect is
opposite: the coupling is increased. Around the physical $\rho $ mass, $%
M=m_{\rho }$, the effective coupling is roughly two times larger than in the
vacuum. For the width of the $\rho \rightarrow \pi \pi $ decay this means a
factor of $4$ enhancement, giving an in-medium width to the $\rho $ of about 
$600$MeV at the saturation density. A similar estimate has been obtained 
{\em e.g.} in Refs. \cite{klingl,eletsky}.

The dot-dashed line in Fig. 2 shows the result of a calculation
with the energy-dependent width, parameterized as a function of the $s$
variable as follows:
\begin{eqnarray}
\Gamma _{\Delta }(s) &=&\Gamma _{\Delta }\left( \frac{q_{{\rm cm}}(s)}{q_{%
{\rm cm}}(m_{\Delta })}\right) ^{3}\theta (s-(m_{N}+m_{\pi })^{2}),
\label{gedep} \\
q_{{\rm cm}}(s) &=&\frac{1}{2\sqrt{s}}\sqrt{(s-(m_{N}+m_{\pi
})^{2})(s-(m_{N}-m_{\pi })^{2})}.
\end{eqnarray}
As can be seen from Fig. 2, the effects of energy-dependent widths are not
large in our analysis.

It is interesting to look at the anatomy of the relative medium contribution
to $g_{{\rm eff}}$, {\em i.e. } of the quantity $\frac{1}{8}\rho _{B}\bar{A}(%
{\bf q}=0)/g_{\rho }$. At $M=m_{\rho }=776$MeV and $\rho _{B}=\rho _{0}=0.17%
{\rm fm}^{-3}$ we find that the diagrams (a)-(g) contribute,
correspondingly, $-0.08$, $0.22-0.09i$, $-0.008-0.008i$, $0.10-0.11i$, $%
-0.11+0.63i$, $-0.008$, and $0.62-1.16i$, with the total of $0.73-0.75i$,
for the case $\Gamma _{\Delta }=120{\rm MeV}$, and $-.08$, $0.33$, $0.001$, $%
0.25$, $-0.87$, $-0.008$, and $2.09$, with the total of $1.71$, for the case 
$\Gamma _{\Delta }=0$. As advocated above, the largest contribution comes
from the $N-\Delta $ bubble diagram (g).

In Fig. 3 we display the results for different values of the baryon density, 
$\rho _{B}$. In this study we have neglected the energy-dependence of the $%
\Delta $ width, but included the density dependence of $\Gamma _{\Delta }$.
The $\Delta $ broadens moderately at the nuclear saturation density \cite
{EW,Oster}, with Pauli blocking giving a decrease of $\Gamma _{\Delta }
$ by about 40MeV, and absorption processes giving an increase by about
80MeV, such that the net effect is an increase by about 40MeV. Therefore we
parameterize 
\begin{equation}
\Gamma _{\Delta }(\rho _{B})=\Gamma _{\Delta }(1+\frac{40{\rm MeV}}{\Gamma
_{\Delta }}\frac{\rho _{B}}{\rho _{0}}) \label{rhopara}.
\end{equation}
Obviously, the effective coupling increases with $\rho _{B}$, as can be seen in Fig. 3.
A caveat is in place here. Our method can be trusted numerically only at low
values of the baryon density, such that the leading-density approximation
holds. On the other hand we can see from Fig. 3 that the effects are large
already at the saturation density, and certainly the approximation breaks
down at larger values of $\rho _{B}$. Therefore our numerical results at
high densities, here and in the following parts of the paper, have to be
taken with a grain of salt and treated as indication of possible large
effects rather than accurate numerical predictions. 

Finally, we complete our discussion of the in-medium $\rho \pi \pi $ vertex
at ${\bf q}=0$ with Fig. 4, which shows the result of the calculation with
fixed $\Gamma _{\Delta }$, but with $m_{N}$ and $m_{\Delta }$ scaled down to
70\% of their vacuum values. A decrease of that order at the saturation
density is anticipated from several approaches \cite{brscale,serot,griegel}.
We notice that $g_{{\rm eff}}$ is enhanced and shifted to lower values of $M$
when the baryon masses are rescaled.

\begin{figure}[tbp]
\vspace{0mm} \epsfxsize = 11 cm \centerline{\epsfbox{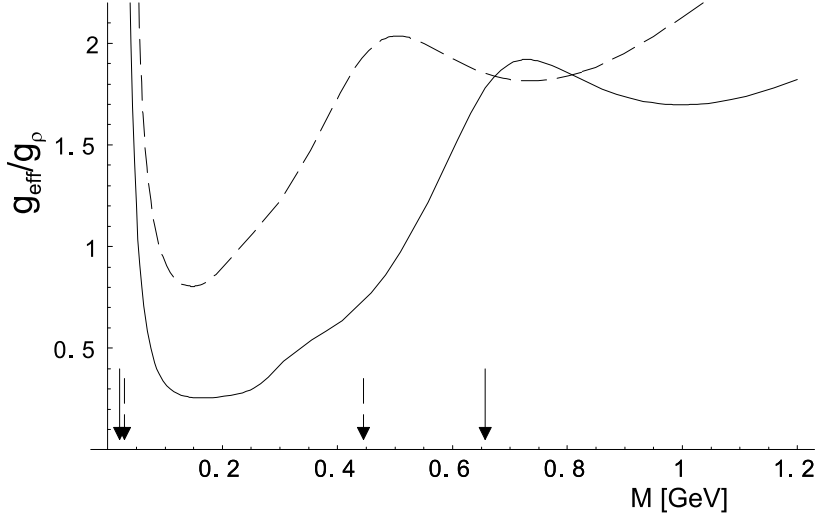}} \vspace{0mm%
} \label{geff07}
\caption{Same as Fig. 2 for the vacuum values of the $m_{N}$ and $m_{\Delta }
$ (solid line), and for the values reduced to 70\% (dashed line). Both
curves evaluated at the baryon saturation density, and for $\Gamma _{\Delta
}=120$MeV.}
\end{figure}

\section{Results for moving $\rho$ \label{moving}}

When the $\rho $ meson moves with a non-zero momentum ${\bf q}$ with respect
to the medium, its propagation is different for transverse and longitudinal
polarizations, defined by quantizing the spin along the direction of ${\bf q}
$. To analyze this effect we shall consider the width of the transversely
and longitudinally polarized $\rho $ mesons due to the decay into two pions.
This allows us to present the result in a more compact form, rather than
looking separately at the functions $\bar{A}$, $\bar{B}$, and $\bar{C}$. The
expression for the width for the decay $\rho^0 \to \pi^+ \pi^-$, as viewed
from the rest frame of the medium, is 
\begin{equation}
\Gamma _{\rho \rightarrow \pi \pi }=\frac{1}{n_{s}}\sum_{s}\frac{1}{2q_{0}}%
\int \frac{d^{3}p}{(2\pi )^{3}2p_{0}}\int \frac{d^{3}p^{\prime }}{(2\pi
)^{3}2p_{0}^{\prime }}|{\cal {M}}|^{2}(2\pi )^{4}\delta ^{(4)}(q-p-p^{\prime
}),  \label{widthg}
\end{equation}
where $n_{s}$ is the number of spin states of the $\rho $ meson, and $%
\sum_{s}$ denotes the sum over these spin states. The division by $q_{0}=%
\sqrt{M^{2}+{\bf q}^{2}}$ in Eq. (\ref{widthg}), rather than by $M$,
accounts for the time-dilatation effect. We perform the phase-space integral
in the rest frame of the nuclear medium, and obtain 
\begin{equation}
\Gamma _{\rho \rightarrow \pi \pi }=\frac{1}{n_{s}}\sum_{s}\frac{1}{2q_{0}}%
\sum_{b=1,2}\int_0^{\gamma^*}\sin \gamma \frac{({\bf p}^{(b)})^{2}}{8\pi
p_{0}^{(b)}(q_{0}-p_{0}^{(b)})|a^{(b)}|}|{\cal {M}}|^{2}d\gamma ,
\label{width}
\end{equation}
where $\sum_{b}$ is the sum over the two possible kinematic branches, which
can appear when the two-body decay is viewed from a frame where the decaying
particle moves. The angle $\gamma $ is between the directions of ${\bf q}$
and ${\bf p}$. The second branch appears only for $|{\bf q|}$ above a
critical value, 
\begin{equation}
|{\bf q|>}\frac{M\sqrt{M^{2}-4m_{\pi }^{2}}}{2m_{\pi }} \equiv q_{\rm crit}. 
\label{qlim}
\end{equation}
Elementary kinematic considerations give 
\begin{eqnarray}
|{\bf p}^{(1,2)}| &=&\frac{M^{2}|{\bf q}|\cos \gamma \pm q_{0}\sqrt{%
M^{4}-4m_{\pi }^{2}(M^{2}+{\bf q}^{2}\sin ^{2}\gamma )}}{2(M^{2}+{\bf q}%
^{2}\sin ^{2}\gamma )},  \nonumber \\
q_{0} &=&\sqrt{M^{2}+{\bf q}^{2}},\,\qquad p_{0}^{(1,2)}=\sqrt{m_{\pi
}^{2}+\left( {\bf p}^{(1,2)}\right) ^{2}},  \nonumber \\
a^{(1,2)} &=&\left. \frac{d(q_{0}-\sqrt{m_{\pi }^{2}+r^{2}}-\sqrt{m_{\pi
}^{2}+r^{2}-2r|{\bf q}|\cos \gamma +{\bf q}^{2}})}{dr}\right| _{r=|{\bf p}%
^{(1,2)}|},  \nonumber \\
\gamma^* &=& \left \{ \begin{array}{l} \pi \hspace{3.8cm} 
{\rm for} \;\;q \le q_{\rm crit} \\ 
\arcsin \left( \frac{M\sqrt{M^{2}-4m_{\pi }^{2}}}{2m_{\pi }|{\bf %
q|}}\right) \hspace{.3cm} {\rm for}\;\; q > q_{\rm crit} 
 \end{array} \right.  .  \label{kinbra}
\end{eqnarray}
The transversely polarized $\rho $ has two helicity states ($n_{s}=2$), with
projection $s=\pm 1$ on the direction of ${\bf q,}$ described by
polarization vectors $\varepsilon _{(\pm )}^{\mu }$, while the
longitudinally polarized $\rho $ has one helicity state ($n_{s}=1$), with
the corresponding projection $s=0$, described by the polarization vector $%
\varepsilon _{(0)}^{\mu }$. An explicit calculation yields \cite{BFH1,disp} 
\begin{eqnarray}
-\varepsilon _{(+)}^{\mu \ast }\varepsilon _{(+)}^{\nu }-\varepsilon
_{(-)}^{\mu \ast }\varepsilon _{(-)}^{\nu } &=&g^{\mu \nu }-u^{\mu }u^{\nu }-%
\frac{(q^{\mu }-q\cdot u\ u^{\mu })(q^{\nu }-q\cdot u\ u^{\nu })}{q\cdot
q-(q\cdot u)^{2}}\equiv T^{\mu \nu },  \nonumber \\
-\varepsilon _{(0)}^{\mu \ast }\varepsilon _{(0)}^{\nu } &=&-\frac{q^{\mu
}q^{\nu }}{q\cdot q}+u^{\mu }u^{\nu }+\frac{(q^{\mu }-q\cdot u\ u^{\mu
})(q^{\nu }-q\cdot u\ u^{\nu })}{q\cdot q-(q\cdot u)^{2}}\equiv L^{\mu \nu }.
\nonumber \\
&&  \label{TL}
\end{eqnarray}
Note that by summing over all polarizations one recovers the usual formula, 
{\em i.e.} 
\begin{equation}
T^{\mu \nu }+L^{\mu \nu }=g^{\mu \nu }-\frac{q^{\mu }q^{\nu }}{q\cdot q}.
\label{sumTL}
\end{equation}
The tensors $T^{\mu \nu }$ and $L^{\mu \nu }$ are defined with such signs as
to form projection operators, {\em i.e. }, $T^{\mu \nu }T_{\nu }^{\cdot
\alpha }=T^{\mu \alpha }$, $L^{\mu \nu }L_{\nu }^{\cdot \alpha }=L^{\mu
\alpha }$, and $T^{\mu \nu }L_{\nu }^{\cdot \alpha }=0$. Furthermore, we
have $T^{\mu \nu }q_{\nu }=0$ and $L^{\mu \nu }q_{\nu }=0$, which reflects
current conservation, as well as $T^{\mu \nu }u_{\nu }=0$. Through the use
of these relation and Eqs. (\ref{Amu},\ref{Amumed}) we find that 
\begin{eqnarray}
&&|{\cal M}_{T}|^{2}=\sum_{s=\pm }\varepsilon _{(s)}^{\mu \ast }A_{\mu
}^{\ast }\,\varepsilon _{(s)}^{\nu }A_{\nu }=-\left| 2(g_{\rho }+\frac{1}{8}%
\rho _{B}\bar{A})\right| ^{2}p_{\mu }T^{\mu \nu }p_{\nu },  \label{widthT} \\
&&|{\cal M}_{L}|^{2}=\sum_{s=0}\varepsilon _{(s)}^{\mu \ast }A_{\mu }^{\ast
}\,\varepsilon _{(s)}^{\nu }A_{\nu }=  \label{widthL} \\
- &&(2(g_{\rho }+\frac{1}{8}\rho _{B}\bar{A})^{\ast }p_{\mu }+\bar{C}^{\ast
}m_{N}u_{\mu })L^{\mu \nu }(2(g_{\rho }+\frac{1}{8}\rho _{B}\bar{A})p_{\nu }+%
\bar{C}m_{N}u_{\nu }).  \nonumber
\end{eqnarray}
Note that the value of the coefficient $\bar{B}$ in Eq. (\ref{Amumed}) is
irrelevant for the widths. Equations (\ref{widthT},\ref{widthL}) are used in
Eq. (\ref{width}).

\begin{figure}[tbp]
\vspace{0mm} \epsfxsize = 11cm \centerline{\epsfbox{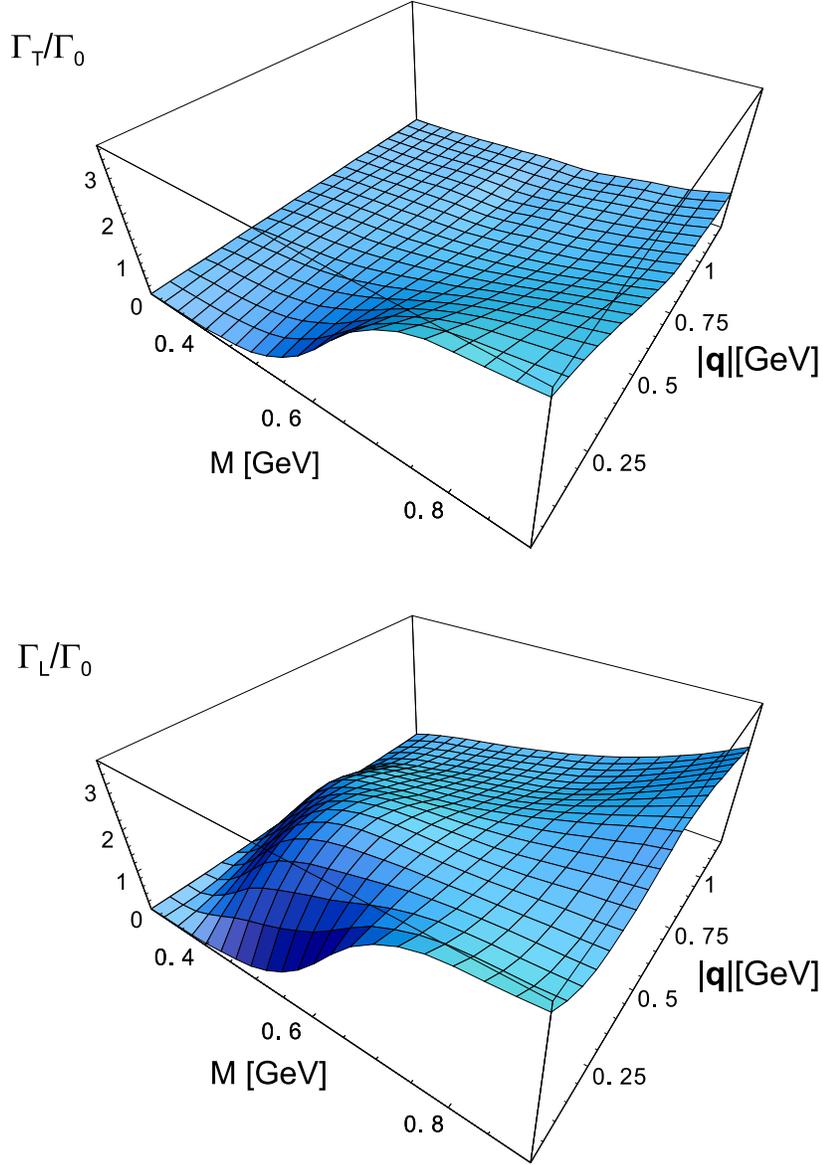}} \vspace{0mm} 
\label{gam3d}
\caption{The ratio of the width for the $\rho \to \pi 
\pi$ decay at the nuclear saturation density to its vacuum value, $%
\Gamma_0$, plotted as a function of the virtual mass of the $\rho$, $%
M$, and the magnitute of its three-momentum with respect to nuclear matter, $%
|{\bf {q}|}$. Top: transverse polarization, bottom: longitudinal
polarization.}
\end{figure}

Our numerical results are shown in Fig. 5. For simplicity, we have
used here the constant $\Gamma _{\Delta }=120{\rm MeV}.$ We notice
considerable dependence on ${\bf q}$, as well as different behavior for the
transverse and longitudinal cases. The transverse width decreases with $|%
{\bf q}|$, while the longitudinal does not. At lower values of $M$ and $|%
{\bf q}|$ around $0.5$GeV the longitudinal width develops a hill, absent in
the transverse case.

The quantity which enters the formula for the dilepton production (see Sect.
\ref{dilep}) is the spectral function of the transverse and longitudinal $%
\rho $ mesons, defined as\footnote{%
The presence of $\sqrt{M^{2}+{\bf q}^{2}}$ here is related to the presence
of $1/q_{0}$ in Eq.(\ref{width}), {\em i.e.} to the fact that we are using
widths viewed from the rest frame of the medium. Had we used widths viewed
from the $\rho $ rest frame, we would have $1/M$ in Eq.(\ref{width}), and
factors of $M$ instead of $\sqrt{M^{2}+{\bf q}^{2}}$ in Eq. (\ref{specstr}).
Of course, in both cases the resulting spectral functions are equal.} 
\begin{equation}
{\cal A}_{P}=\frac{1}{\pi }\frac{\sqrt{M^{2}+{\bf q}^{2}}\Gamma _{P}}{%
(M^{2}-m_{\rho }^{2})^{2}+(M^{2}+{\bf q}^{2})\Gamma _{P}^{2}},\qquad
P=T,L,  \label{specstr}
\end{equation}
where $m_{\rho }$ is the position of the pole. The results for $%
{\cal A}_{T}$ and ${\cal A}_{L}$, with $m_{\rho }=776$MeV,
are plotted in Fig. 6. We note that while at ${\bf q}=0$ we obviously have $%
{\cal A}_{T}={\cal A}_{L}$, at larger values of ${\bf q}$ and at $M$ around $%
m_{\rho }$ the transverse spectral strength becomes dominant. At first
glance this may seem surprising, since in Fig. 5 we have seen that at higher 
${\bf q}$ we have much larger $\Gamma _{L}$ than $\Gamma _{T}$. However, the
optimum value of $\Gamma _{P}$, at which ${\cal A}_{P}$ has a maximum, is $%
\Gamma _{P}=(M^{2}-m_{\rho }^{2})/\sqrt{M^{2}+{\bf q}^{2}}$. Lower, as
well as higher values of $\Gamma _{P}$ lead to a decrease in ${\cal A}_{P}$.
This explains the behavior of Fig. 6. We note that the transverse spectral
strength, ${\cal A}_{T}$, is concentrated along a ridge extending far into
the large-${\bf q}$ region. Thus, a proper description of propagation at
finite and large values of ${\bf q}$ is needed for the description of $\rho $
mesons in medium. We note that our results are in qualitative agreement with
Ref. \cite{Urban0} (Fig. 14), with the somewhat different behavior of ${\cal A%
}_{T}$, which reaches larger values at higher values of $|{\bf q}|$ in our
approach.\footnote{Note a factor of $1/\pi $ difference in our
definition of the spectral functions compared to those of Ref. \cite{Urban0}.}
We also find qualitative similarity to the results of the
altogether different model of Ref. \cite{Peters} (Figs. 6,7).
There is a difference at 
larger values of $|{\bf q}|$  for ${\cal A}_T$, 
manifest in the presence of the rim in our Fig. 6. 
The results for  ${\cal A}_L$
are very similar to \cite{Peters}. 

\begin{figure}[tbp]
\vspace{0mm} \epsfxsize = 11cm \centerline{\epsfbox{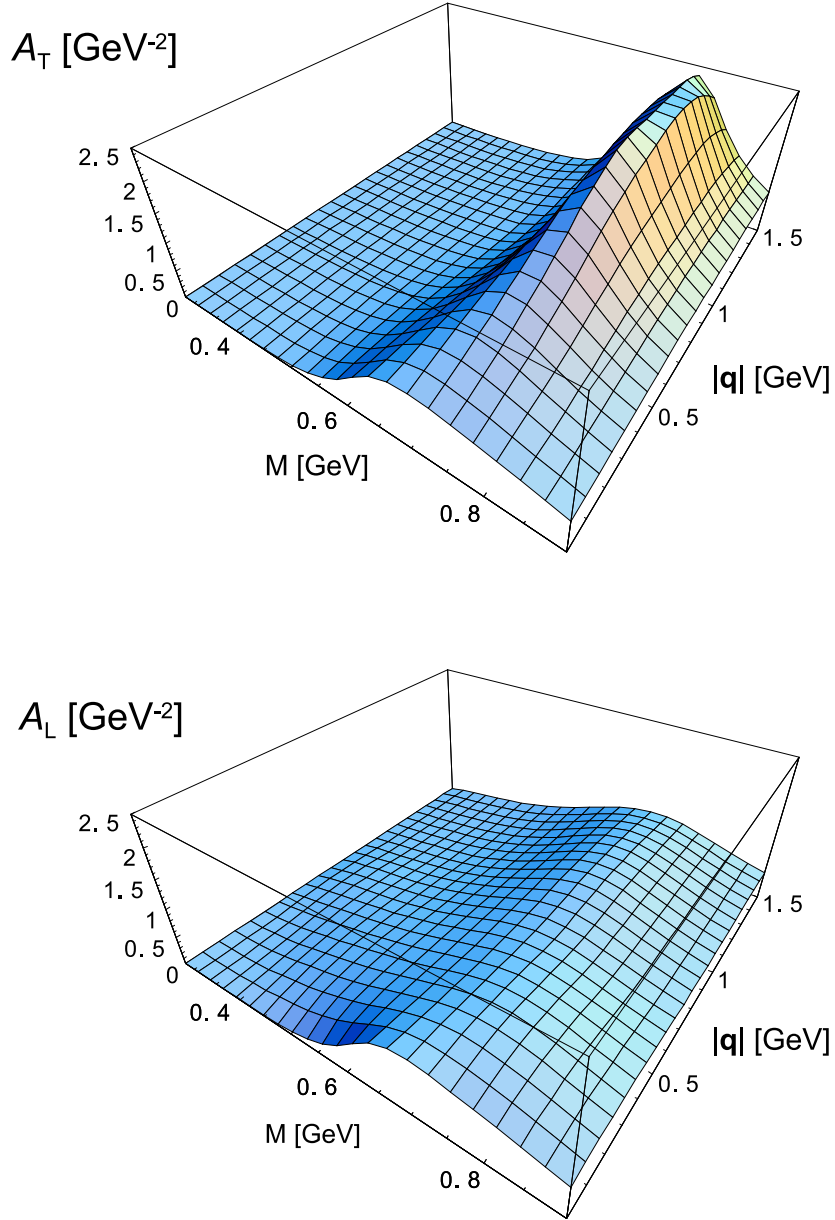}} \vspace{0mm} 
\label{a3d}
\caption{The spectral strengths in the $\rho $ channel at the
nuclear saturation density, corresponding to the widths of Fig. 5. Top:
transverse polarization, bottom: longitudinal polarization. }
\end{figure}

We should stress here that our construction of the $\rho $ spectral function
accounts only for the process $\rho \rightarrow \pi \pi $ and ignores all
other possible contributions, such as {\em e.g.} from 
the $s$-channel resonances,
studied in Refs. \cite{Post,Peters}, medium modifications of the 
pions, {\em etc.} Such processes 
should be included in a more complete calculation. 
Still, the results of Sect. \ref{dilep} depend only on the shape of the
spectral functions of Fig. 6, and not on the physics 
leading to their form. Since the spectral functions obtained in 
many other approaches are quite similar to ours,  
the results obtained below 
can be viewed as representative to approaches containing the broadening of the 
$\rho$.

\section{Dilepton production rate\label{dilep}}

Measurements of the low-mass dilepton spectra \cite{ceres,helios} have shown
significant excess above yields from the final-state hadron decays. In this
context, the properties of vector mesons (especially of the $\rho $) in a
hadronic environment become of particular interest, since the Vector Meson
Dominance Model is commonly used to make the estimates of the dilepton
yields from vector-meson decays. There are numerous analysis of the effect
in the literature, both in hydrodynamic approaches \cite
{Dom,Sriva,HungDil,Sollfrank,BaierDil}, and in transport theories
\cite{rapp,KochDil,li,LiDil,CassingDil,RappDil,SchulzeDil,SteeleDil}. The
dilepton-rate formula \cite{GaleKap,GaleKap2,KorpaPr,Weldon,Pkoch} from $%
\rho $ meson decays can be written in a manifestly Lorentz covariant way as
follows: 
\begin{equation}
{\frac{dN}{d^{4}x\,dM^{2}}}=\int {\frac{d^{3}q}{(2\pi )^{3}}}{\frac{M}{E_{q}}%
}\Gamma _{\rho \rightarrow e^{+}e^{-}}{\cal A}\left( M,q\cdot u,\rho
_{B}(x)\right) f_{\rho }\left( {\frac{q\cdot u}{T(x)}}\right) ,  \label{dN}
\end{equation}
where $M$ is the invariant mass of the lepton pair, equal to the mass of the
virtual $\rho $ meson, ${\cal A}=2{\cal A}_{T}+{\cal A}_{L}$ is the spectral
function including all polarizations, $x$ is a space-time point, $T(x)$ is
the local value of temperature, and $E_{q}=\sqrt{M^{2}+{\bf q}^{2}}$. The
quantity $\Gamma _{\rho \rightarrow e^{+}e^{-}}$ in Eq. (\ref{dN}) is the
width for the process $\rho \rightarrow e^{+}e^{-}$, 
\begin{equation}
\Gamma _{\rho \rightarrow e^{+}e^{-}}={\frac{4\pi \alpha _{QED}^{2}m_{\rho
}^{4}}{3g_{\rho }^{2}M^{5}}}\left( 1-{\frac{4m_{e}^{2}}{M^{2}}}\right) ^{{%
\frac{1}{2}}}\left( M^{2}+2m_{e}^{2}\right) ,  \label{Grhodil}
\end{equation}
where $\alpha _{QED}$ is the fine structure constant, and $m_{e}$ is the
mass of the electron. Finally, the function $f_{\rho }$ in Eq. (\ref{dN}) is
the thermal Bose-Einstein distribution of the $\rho $ mesons, 
\begin{equation}
f_{\rho }=\left[ \hbox{exp}\left( {\frac{q\cdot u-2\mu _{\pi }}{T}}\right) -1%
\right] ^{-1},  \label{be}
\end{equation}
whith $\mu _{\pi }$ denoting the pion chemical potential \cite
{Pkoch,KochChem}, incorporated in several works. This quantity in some sense
mimics possible deviations of the system from the chemical equilibrium.

In order to describe the problem as realistically as possible, we will
include the effects of the {\em expansion} of the medium formed in a
relativistic heavy-ion collision. The lepton pairs are formed in a fire
cylinder which moves as a whole in the lab system with the rapidity $\alpha
_{FC}$. For symmetric and central collisions $\alpha _{FC}$ is a half of the
projectile rapidity in the lab. In its own center-of-mass system (CM), the
fire cylinder undergoes a {\em hydrodynamic expansion}. In the analysis of
such a situation, it is convenient to rewrite Eq. (\ref{dN}) in the
variables suited to both the kinematics of the emission process and the
geometry of the experimental setup. We introduce $M_{\perp }=\sqrt{%
M^{2}+q_{\perp }^{2}}$, the transverse mass of the dilepton pair, $y^{lab}$,
the rapidity of the pair measured in the lab system, ${\bf u}_{\perp }$, the
transverse four-velocity of the fluid element producing dileptons, and $%
\alpha ^{lab}$, the rapidity of this fluid element in the lab. With these
variables we have 
\begin{equation}
q\cdot u=M_{\perp }\sqrt{1+u_{\perp }^{2}}\hbox{cosh }(y^{lab}-\alpha
^{lab})-{\bf q}_{\perp }\cdot {\bf u}_{\perp }.  \label{qu}
\end{equation}
The velocity of the fluid element in the lab is a relativistic superposition
of the velocity of the fire cylinder in the lab and the hydrodynamic flow
considered in the CM system. Thus we have 
\begin{equation}
\alpha ^{lab}=\alpha +\alpha _{FC}=\hbox{arctanh }v_{||}+\alpha _{FC},%
\hspace{0.5cm}u_{\perp }={\frac{v_{\perp }\hbox{cosh}(\alpha )}{\sqrt{%
1-v_{\perp }^{2}\hbox{cosh}^{2}(\alpha )}}}.  \label{alabuperp}
\end{equation}
We note that the velocities $v_{||}$ and $v_{\perp }$ are defined now in the
CM system. They depend on time and space coordinates.

Next, we analyze the kinematic constraint of the CERES experiment. The
experimental acceptance cuts can be included with help of the function 
\begin{eqnarray}
&&\Phi (M,y^{lab},q_{\perp })={\frac{\int d^{2}p_{1\perp }d^{2}p_{2\perp
}dy_{1}dy_{2}\,\,\,{\bf \phi }\,\,\,\delta (E_{q}-E_{p_{1}}-E_{p_{2}})\delta
^{(3)}({\bf q}-{\bf p}_{1}-{\bf p}_{2})}{\int d^{2}p_{1\perp }d^{2}p_{2\perp
}dy_{1}dy_{2}\delta (E_{q}-E_{p_{1}}-E_{p_{2}})\delta ^{(3)}({\bf q}-{\bf p}%
_{1}-{\bf p}_{2})}},  \nonumber \\
&&  \label{Phi}
\end{eqnarray}
where ${\bf p}_{1,\,2}$ are the momenta of the emitted electrons, $y_{1,\,2}$
are the electron rapidities, and ${\bf \phi }$ is a product of step
functions which enforces the experimental setup conditions: $2.1=\eta
_{min}<y_{1,\,2}<\eta _{max}=2.65$, $p_{1,\,2}^{\,\perp }>$ 200 MeV, and $%
\theta _{ee}>$ 35mrad. Due to the smallness of the electron mass, we can
assume here that rapidities and pseudorapidities of the electrons are equal.
The construction of the function $\Phi (M,y^{lab},q_{\perp })$ 
requires a numerical calculation of a two-dimensional integral of 
a function involving a product of step functions, which is very 
easily accomplished by a Monte Carlo method.  
With the inclusion of the experimental acceptance cuts, the dilepton
production rate is 
\begin{equation}
{\frac{dN}{d^{4}x\,dM\Delta \eta }}={\frac{2M^{2}}{(\eta _{max}-\eta _{min})}%
}\int {\frac{d^{2}q_{\perp }}{(2\pi )^{3}}}\int dy^{lab}\,\Phi \,\Gamma
_{\rho \rightarrow e^{+}e^{-}}\,{\cal A}\,f_{\rho }.  \label{dN1}
\end{equation}
One should stress here the relevance of the inclusion of the kinematic cuts
for the obtained results. The function $\Phi $ influences mostly the overall
normalization of the cross section, and not so much the dependence on $M$.

In order to calculate the dilepton spectrum, one has to assume a model of
the hydrodynamic expansion of the fire cylinder. We adopt the fire-cylinder
expansion model of Refs. \cite{Rappevol,RappShur}. It is assumed that the
system is in thermal equilibrium up to time $t_{max}$, when freeze-out
occurs, and the velocities depend on space-time in the following way: 
\begin{equation}
v_{||}(t,z)=(v_{z}+a_{z}t)\frac{z}{z_{max}(t)},\qquad v_{\perp
}(t,r)=(v_{r}+a_{r}t)\frac{r}{r_{max}(t)},
\end{equation}
where 
\begin{equation}
z_{max}(t)=z_{0}+v_{z}t+\frac{1}{2}a_{z}t^{2},\qquad r_{max}(t)=r_{0}+v_{r}t+%
\frac{1}{2}a_{r}t^{2}
\end{equation}
are the boundaries of the system at time $t$. The parameters of the
expansion are as follows  \cite{Rappevol,RappShur}: 
\begin{eqnarray}
t_{max} &=&11{\rm fm},\quad z_{0}=4.55{\rm fm,\quad }r_{0}=4.6{\rm fm,} \\
v_{z} &=&0.5,\quad a_{z}=0.023{\rm fm}^{-1},\quad v_{r}=0,\quad a_{r}=0.05%
{\rm fm}^{-1},
\end{eqnarray}
and the time dependences of the temperature and the baryon density are: 
\begin{equation}
T(t)=210{\rm MeV}\exp \left( -\frac{t}{18.26{\rm fm}}\right) ,\qquad \rho
_{B}(t)=260/V(t),
\end{equation}
where $V(t)=2\pi z_{max}(t)r_{max}^{2}(t)$ is the volume of the fire
cylinder at time $t$. For the time dependence of the pion chemical
potential, $\mu _{\pi }(t)$, we assume a linear rise from $20{\rm MeV}$ at $%
t=0$ to $80{\rm MeV}$ at $t=t_{max}$ \cite{rapp}.

Finally, the yield of leptons produced during the expansion is 
\begin{equation}
{\frac{dN_{1}}{dM\Delta \eta }}=\int\limits_{0}^{t_{max}}dt\int%
\limits_{0}^{r_{max}(t)}2\pi
rdr\int\limits_{-z_{max}(t)}^{z_{max}(t)}dz\left( {\frac{dN}{%
d^{4}x\,dM\Delta \eta }}\right) ,  \label{expyield}
\end{equation}
where ${dN/(d^{4}x\,dM\Delta \eta )}$ is given by Eq. (\ref{dN1}) with all
the required substitutions.

In addition to the yields of Eq. (\ref{expyield}) one usually adds the
contribution from vector mesons which remain after freeze-out. This
contribution is equal to 
\begin{equation}
{\frac{dN_{2}}{dM\Delta \eta }}=\frac{1}{\Gamma (M)}\int%
\limits_{0}^{r_{max}(t_{max})}2\pi
rdr\int\limits_{-z_{max}(t_{max})}^{z_{max}(t_{max})}dz\left( {\frac{dN}{%
d^{4}x\,dM\Delta \eta }}\right) ,  \label{tailyield}
\end{equation}
where $\Gamma (M)$ is the {\em full} width of the $\rho $ meson with virtual
mass $M$, given by the formula 
\begin{equation}
\Gamma (M)=\frac{g_{\rho \pi \pi }^{2}}{48\pi M^{2}}(M^{2}-4m_{\pi
}^{2})^{3/2}
\end{equation}
with ${g}_{\rho \pi \pi }=5.98$ giving $\Gamma (m_{\rho })=150$MeV. The
physical interpretation of formula (\ref{tailyield}) is that all $\rho $
mesons that remain after freeze-out, decay with the yield proportional to
the number of mesons and the branching ratio to the dilepton channel.
Finally, the full contribution is 
\begin{equation}
{\frac{dN}{dM\Delta \eta }}={\frac{dN_{1}}{dM\Delta \eta }}+{\frac{dN_{2}}{%
dM\Delta \eta }}.  \label{lepyield}
\end{equation}

\begin{figure}[tbp]
\vspace{0mm} \epsfxsize = 12.5 cm \centerline{\epsfbox{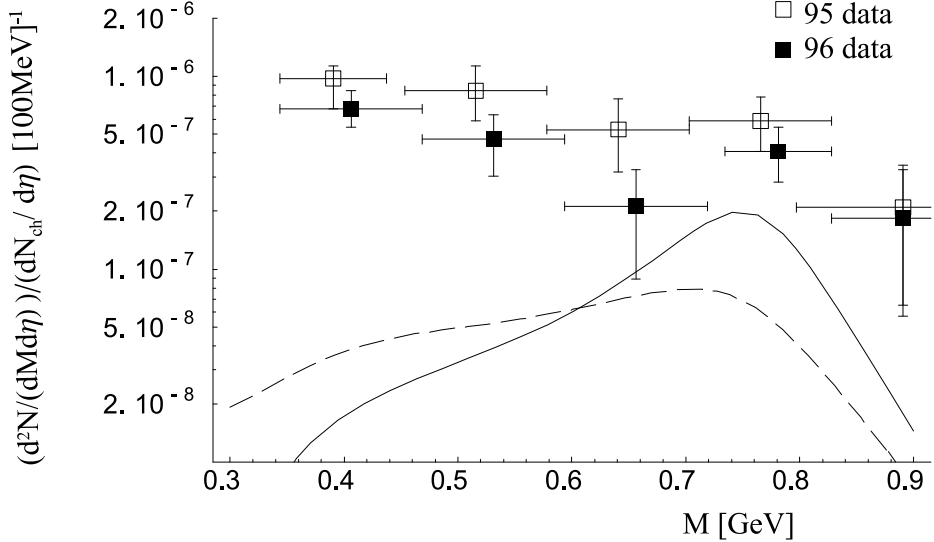}} \vspace{%
0mm} \label{ceres}
\caption{Dilepton yields for the 158GeV/A $Pb+Au$ CERES experiment from the $%
\rho$ decays. The solid line is the result of the calculation with
Eq. (\protect\ref{expyield}) and the vacuum $\rho$ spectral function. The
dashed line is obtained with the medium-modified 
spectral strength, Eq. (\protect\ref
{specstr}). Contributions from other processes or the ``cocktail''
background are not included in the theoretical curves.}
\end{figure}

Our numerical results are shown in Fig. 7. The solid line shows the yield
from the $\rho $ decays, Eq. (\ref{lepyield}), with the vacuum spectral
strength of the $\rho $, while the dashed curve shows the calculation with
our medium-modified spectral strength of Eq. (\ref{specstr}). We can see
that the medium effects redistribute the dilepton yields from higher to
lower values of $M$. This is a typical effect of broadening of the $\rho ,$
found in previous investigations. Thus the tendency needed to explain the
experimental data is correct. We can see by comparing the dashed line to the
data that the calculated yield falls an order of magnitude below the data.
Note that other processes, not included in our calculation, contribute in
the explored region of $M$, for instance the Dalitz decays of mesons, or the 
$\omega $ decays. Also, note that we are not including the ``cocktail''
contribution of decays of hadrons in our comparison, which would not be
consistent. We have found that 15-25\% of the model yields comes from decays
after freeze-out. The effects of the expansion of the fire cylinder enhance
the yields by a few percent. We stress that the overall normalization of the
calculated curve is sensitive to the time-integrated volume of the fire
cylinder, and to the hydrodynamic expansion parameters \cite{HungFire}.

\section{Summary and discussion\label{summa}}

Here are the main conclusions of our investigation: the medium
effects on the $\rho \pi \pi $ coupling are large, and dominantly come from
the process where the $\Delta $ is excited in the intermediate state. At the
nuclear saturation density and at physical mass of the ${\rho }$ 
the value of the
coupling is roughly doubled compared to the vacuum value. The increased
coupling leads directly to large widths of the $\rho $ meson in medium. We
have analyzed the resulting spectral functions for the transverse and
longitudinal polarizations, with the result that at higher values of the
three-momentum with respect to the medium, the transverse spectral function
is much larger from the longitudinal one. Finally, we have applied our model
to evaluate the dilepton production from the $\rho $ decays in relativistic
heavy-ion collisions. We confirm the well-known finding that a larger $\rho $
width helps to understand the experimental data by redistributing strength
from higher to lower invariant dilepton masses. However, the overall norm of
the dilepton yields from the $\rho $ decays is almost an order of magnitude
too small to explain the experimental data. We remark that the results for
the dilepton yields from $\rho $ decays are only sensitive to the actual
shape of the spectral functions and to the hydrodynamic expansion. In that
sense they are not sensitive to detailed modelling of the dynamics, as long
as the resulting functions ${\cal A}_{T}$ and ${\cal A}_{L}$ are similar.
Thus our results confirm the statement of Ref. \cite{HungFire}, namely that
hydrodynamic models have problems in explaining the dilepton data unless the
hydrodynamic evolution is exceedingly long.

For simplicity, we have used on-shell hadronic couplings throughout our
studies. Since the nucleon and the $\Delta $ in the hadronic loop of Fig. 1
can be off-shell, additional coupling structures may be present. In
addition, sideways form factors could be included for the particles
off-shell. Presently, this has not been done, again for simplicity and from
the lack of knowledge as to how to introduce and choose these form factors.
One could also include the medium modifications of the meson-baryon
couplings, a feature advocated, {\em e.g.}, in the model of Ref. \cite
{Banerjee92,Banerjee96}.

The width of the $\rho $ meson picks up contributions not only from the
pion-loop diagram, included in our work, but also from other processes. In
particular, one can include the $s$-channel resonances, as studied {\em e.g.}
in Refs. \cite{Post,Peters}. Such processes can and should be included in a
more complete and realistic calculation. Also the diagrams of Fig. 1 could
in principle be supplied with higher resonances, at the expense of having
more not well known parameters. 

In addition to the effects of the Fermi sea, studied in this paper, {\em %
vacuum polarization} effects may influence the $\rho \pi \pi $ coupling. To
have an estimate of these effects, we have done a Walecka-type calculation
where in the diagrams of Fig. 1 we have included the {\em free} nucleon
propagators only. We have found a $\sim $10\% increase of the coupling when
the nucleon mass is reduced from 939MeV to 700MeV. We have applied
Pauli-Villars regulator with the cut-off parameter of 1GeV in order to
truncate high momenta in the loop. Similar order of the effect is found when
the Nambu--Jona-Lasinio model is used along the lines of Ref. \cite{Bernard}%
, and the mass of the quark is scaled down as expected from the medium
effects. Thus, vacuum polarization effects on the $\rho \pi \pi $ coupling
are estimated to be less significant than the Fermi-sea effects presented in
this paper.

Concerning the methods applied,
we have shown in detail how to obtain 
very simply the leading-density approximation for diagrams 
involving loops with density-dependent nucleon propagators. We have also
demonstrated how to carry a Lorentz-covariant calculation of the 
transverse and longitudinal spectral functions with help of the 
formulas of Sect. \ref{moving}. These methods and formulas, although
very straightforward, are not, to our knowledge, commonly know. They
can be useful in studies similar to ours.

\bigskip One of us (WF) thanks Dariusz Mi\'skowiec for a discussion of the
CERES experimental cuts.

\appendix

\section{Rarita-Schwinger spinors\label{rar}}

The Rarita-Schwinger spinors are defined as 
\begin{equation}
u^{\mu }(p,s_{\Delta })=\sum_{\lambda ,s_{N}}\langle 1\frac{1}{2}\lambda s|1%
\frac{1}{2}\frac{3}{2}s_{\Delta }\rangle e^{\mu }(p,\lambda )u_{\Delta
}(p,s),  \label{umuRS}
\end{equation}
with 
\begin{eqnarray}
e^{0}(p,\lambda ) &=&\frac{\vec{\varepsilon}_{\lambda }\cdot \vec{p}}{%
m_{\Delta }},\qquad e^{i}(p,\lambda )=\varepsilon^i_{\lambda }+\frac{(\vec{%
\varepsilon}_{\lambda }\cdot \vec{p})p^{i}}{m_{\Delta }(E_{\Delta
}+m_{\Delta })},\qquad  \label{ED} \\
E_{\Delta } &=&\sqrt{\vec{p}^{2}+m_{\Delta }^{2}}.
\end{eqnarray}
The polarization vectors are defined as 
\begin{equation}
\vec{\varepsilon}_{0}=\left( 
\begin{array}{c}
0 \\ 
0 \\ 
1
\end{array}
\right) ,\qquad \vec{\varepsilon}_{\pm }=\frac{1}{\sqrt{2}}\left( 
\begin{array}{c}
\mp 1 \\ 
-i \\ 
0
\end{array}
\right) ,  \label{e0pm}
\end{equation}
and 
\begin{equation}
u_{\Delta }(p,s)=\sqrt{\frac{E_{\Delta }+m_{\Delta }}{2m_{\Delta }}}\left( 
\begin{array}{c}
1 \\ 
\frac{\vec{\sigma}\cdot \vec{p}}{E_{\Delta }+m_{\Delta }}
\end{array}
\right) \chi (s),  \label{uD}
\end{equation}
where $\chi (s)$ is the two-component spinor. The spinor $u^{\mu
}(p,s_{\Delta })$ satisfies the conditions $\gamma _{\mu }u^{\mu
}(p,s_{\Delta })=0$ and $p_{\mu }u^{\mu }(p,s_{\Delta })=0$.

\section{Isospin algebra\label{isoalg}}

The isospin ${\frac{1}{2}}\rightarrow {\frac{3}{2}}$ transition matrices are
defined through the Clebsch-Gordan coefficients as follows: $\langle \frac{3%
}{2},I_{3}|T^{\mu }|\frac{1}{2},i_{3}\rangle =\langle \frac{1}{2}1i_{3}\mu |1%
\frac{1}{2}\frac{3}{2}I_{3}\rangle $, with $i_{3}$ and $I_{3}$ denoting the
isospin of the nucleon and $\Delta $, respectively. In Cartesian basis the
explicit form reads 
\begin{equation}
T^{1}=\left( 
\begin{array}{cc}
-\frac{1}{\sqrt{2}} & 0 \\ 
0 & -\frac{1}{\sqrt{6}} \\ 
\frac{1}{\sqrt{6}} & 0 \\ 
0 & \frac{1}{\sqrt{2}}
\end{array}
\right) ,\qquad T^{2}=i\left( 
\begin{array}{cc}
\frac{1}{\sqrt{2}} & 0 \\ 
0 & \frac{1}{\sqrt{6}} \\ 
\frac{1}{\sqrt{6}} & 0 \\ 
0 & \frac{1}{\sqrt{2}}
\end{array}
\right) ,\qquad T^{3}=\left( 
\begin{array}{cc}
0 & 0 \\ 
\sqrt{\frac{2}{3}} & 0 \\ 
0 & \sqrt{\frac{2}{3}} \\ 
0 & 0
\end{array}
\right) .  \label{Tmat}
\end{equation}
where the columns are labeled by $i_{3}=\frac{1}{2},-\frac{1}{2}$, left to
right, and the rows by $I_{3}=\frac{3}{2},\frac{1}{2},-\frac{1}{2},-\frac{3}{%
2}$, top to bottom. The following useful relation holds:
\begin{equation}
T^{a\dagger }T^{b}=\frac{2}{3}\delta ^{ab}-\frac{1}{3}\varepsilon ^{abc}\tau
_{c}.  \label{Talg}
\end{equation}
The couplings of the $\Delta $ to the isovector proceed via the matrix $%
T_{\Delta }^{\mu }$ defined as $\langle \frac{3}{2},I_{3}^{\prime
}|T_{\Delta }^{\mu }|\frac{3}{2},I_{3}\rangle =\frac{\sqrt{15}}{2}\langle 
\frac{3}{2}1I_{3}^{\prime }\mu |\frac{3}{2}\frac{1}{2}\frac{3}{2}%
I_{3}\rangle $. Explicitly, we find 
\begin{eqnarray}
T_{\Delta }^{1} &=&\left( 
\begin{array}{cccc}
0 & \frac{\sqrt{3}}{2} & 0 & 0 \\ 
\frac{\sqrt{3}}{2} & 0 & 1 & 0 \\ 
0 & 1 & 0 & \frac{\sqrt{3}}{2} \\ 
0 & 0 & \frac{\sqrt{3}}{2} & 0
\end{array}
\right) ,\qquad T_{\Delta }^{2}=i\left( 
\begin{array}{cccc}
0 & -\frac{\sqrt{3}}{2} & 0 & 0 \\
\frac{\sqrt{3}}{2} & 0 & -1 & 0 \\ 
0 & 1 & 0 & -\frac{\sqrt{3}}{2} \\ 
0 & 0 & \frac{\sqrt{3}}{2} & 0
\end{array}
\right) ,\qquad  \nonumber \\
T_{\Delta }^{3} &=&\left( 
\begin{array}{cccc}
\frac{3}{2} & 0 & 0 & 0 \\ 
0 & \frac{1}{2} & 0 & 0 \\ 
0 & 0 & -\frac{1}{2} & 0 \\ 
0 & 0 & 0 & -\frac{3}{2}
\end{array}
\right) ,  \label{TDelta}
\end{eqnarray}
where the columns are labeled by $I_{3}^{\prime }=\frac{3}{2},\frac{1}{2},-%
\frac{1}{2},-\frac{3}{2}$, left to right, and the rows by $I_{3}=\frac{3}{2},%
\frac{1}{2},-\frac{1}{2},-\frac{3}{2}$, top to bottom. The conventional
factor of $\sqrt{15}/2$ in the definition ensures that $T_{\Delta }^{3}$
simply measures the third component of the isospin of the $\Delta .$ The
spin coupling matrices used in non-relativistic calculations, $S^{i}$ and $%
S_{\Delta }^{i}$, are defined analogously and have exactly the same values
as $T^{a}$ and $T_{\Delta }^{a}$.

Isospin trace factors for diagrams of Fig. 1 can be now readily obtained.
They are equal to $2$, $-\frac{2}{3}$, $-\frac{2}{3}$, $\frac{5}{3}$, $\frac{%
5}{3}$, $2$, and $\frac{4}{3}$ times $-i\epsilon ^{acb}$, for diagrams
(a),(b),...,(g), respectively. Isospin indices $a$, $b$, and $c$ have the
assignment as specified on the text above Eq. (\ref{rppvac}).

\section{Amplitudes for ${\bf q}=0$}
\label{ampliexp}

For the case ${\bf q}=0$ the amplitude corresponding to the diagram $(i)$ of
Fig. 1 can be written as 
\begin{equation}
A_{(i)}^{\mu }=\frac{1}{8}\rho _{B}G_{(i)}\frac{N_{(i)}}{D_{(i)}}(2p^{\mu
}-q^{\mu }),
\end{equation}
where $G_{(i)}$ are products of coupling constants: 
\begin{eqnarray}
G_{(a)} &=&\frac{g_{A}^{2}g_{\rho }}{8F_{\pi }^{2}},\qquad G_{(b)}=\frac{%
\sqrt{2}g_{A}f^{\ast }f_{\pi N\Delta }}{2F_{\pi }m_{\pi }^{2}},\qquad
G_{(c)}=\frac{g_{\rho }f_{\pi N\Delta }^{2}}{2m_{\pi }^{2}},  \nonumber \\
G_{(d)} &=&\frac{g_{\rho }f_{\pi N\Delta }^{\,\,2}}{m_{\pi }^{2}},\qquad
G_{(e)}=\frac{3\sqrt{2}f^{\ast }f_{\pi N\Delta }f_{\Delta }}{2m_{\pi }^{3}}%
,\qquad G_{(f)}=\frac{g_{\rho }g_{A}^2}{4F_{\pi}^2},  \nonumber \\
G_{(g)} &=&\frac{g_{\rho}f_{\pi N\Delta }^2}{m_{\pi }^{2}}.
\end{eqnarray}

The formulas for $N_{(i)}$ and $D_{(i)}$ are very long in the general case,
which reflects the presence of many terms in the Rarita-Schwinger
propagator. Numerators $N_{(i)}$ become manageable in the formal case $%
\Gamma _{\Delta }=0$, and $m_{\pi }=0$, where we find 
\begin{eqnarray}
\hspace{-5mm}N_{(a)} &=&-8\,\left( 4\,m_{N}^{4}\,{M}^{2}+{\kappa _{\rho }}%
\,m_{N}^{2}\,{M}^{4}\right) ,  \nonumber \\
\hspace{-5mm}N_{(b)} &=&16\,m_{N}^{2}\,\left( m_{N}+{m_{\Delta }}-{M}\right)
\,{M}^{2}\,\left( m_{N}+{m_{\Delta }}+{M}\ \right) \,  \nonumber \\
&&\left( 2\,m_{N}^{4}+m_{N}^{3}\,{m_{\Delta }}-m_{N}\,{{m_{\Delta }}}^{3}+2\,%
{{m_{\Delta }}}^{4}-2\,m_{N}^{2}\,\left( 2\,{{m_{\Delta }}}^{2}+{M}%
^{2}\right) \right) ,  \nonumber \\
\hspace{-5mm}N_{(c)} &=&-16\,m_{N}\,\left( m_{N}-{m_{\Delta }}\right)
\,\left( m_{N}+{m_{\Delta }}\right) \,  \nonumber \\
&&\left( m_{N}^{3}-\left( 1+{\kappa _{\rho }}\right) \,m_{N}^{2}\,{m_{\Delta
}}-2\,m_{N}\,{{m_{\Delta }}}^{2}+\left( 1+{\kappa _{\rho }}\right) \,{{%
m_{\Delta }}}^{3}\right) \,{M}^{2}-  \nonumber \\
&&4\,\left(
\begin{array}{c}
\left( -4+{\kappa _{\rho }}\ \right) \,m_{N}^{4}+4\,\left( 1+{\kappa _{\rho }%
}\right) \,m_{N}^{3}\,{m_{\Delta }}- \\ 
\left( 2+5\,{\kappa _{\rho }}\right) \,m_{N}^{2}\,{{m_{\Delta }}}^{2}+2\,{%
\kappa _{\rho }}\,{{m_{\Delta }}}^{4}\ 
\end{array}
\right) \,{M}^{4}+4\,{\kappa _{\rho }}\,m_{N}^{2}\,{M}^{6},  \nonumber \\
\hspace{-5mm}N_{(d)} &=&-20\,{M}^{2}\,\left( 4\,m_{N}^{3}\,{m_{\Delta }}%
-m_{N}\,{{m_{\Delta }}}^{3}-4\,{{m_{\Delta }}}^{4}+m_{N}^{2}\,\left( 7\,{{%
m_{\Delta }}}^{2}+{M}^{2}\right) \ \right) ,  \nonumber \\
\hspace{-5mm}N_{(e)} &=&40\,{M}^{2}\,\left(
\begin{array}{c}
\left( m_{N}-{m_{\Delta }}\right) \,{\left( m_{N}+{m_{\Delta }}\right) }%
^{3}\,\left(
\begin{array}{c}
2\,m_{N}^{4}+4\,m_{N}^{3}\,{m_{\Delta }}- \\
2\,m_{N}\,{{m_{\Delta }}}^{3}-7\,{{m_{\Delta }}}^{4}
\end{array}
\right) - \\
\left( m_{N}+{m_{\Delta }}\right) \,\left(
\begin{array}{c}
3\,m_{N}^{5}+7\,m_{N}^{4}\,{m_{\Delta }}+4\,m_{N}^{3}\,{{m_{\Delta }}}^{2}+
\\
m_{N}^{2}\,{{m_{\Delta }}}^{3}-2\,m_{N}\,{{m_{\Delta }}}^{4}+2\,{{m_{\Delta }%
}}^{5}
\end{array}
\right) \,{M}^{2}+ \\
{m_{\Delta }}\,\left( m_{N}+{m_{\Delta }}\right) \,\left(
2\,m_{N}^{2}+m_{N}\,{m_{\Delta }}-2\,{{m_{\Delta }}}^{2}\right) \,{M}%
^{4}+m_{N}^{2}\,{M}^{6}
\end{array}
\right) ,  \nonumber \\
\hspace{-5mm}N_{(f)} &=&0,  \nonumber \\
\hspace{-5mm}N_{(g)} &=&16\,\left(
\begin{array}{c}
4\,\left( m_{N}-{m_{\Delta }}\right) \,{{m_{\Delta }}}^{2}\,{\left( m_{N}+{%
m_{\Delta }}\right) }^{2}+ \\ 
m_{N}\,\left( m_{N}^{2}+2\,m_{N}\,{m_{\Delta }}-{{m_{\Delta }}}^{2}\right) \,%
{M}^{2}
\end{array}
\right) .
\end{eqnarray}

The expressions for $D_{(i)}$ for the case $\Gamma _{\Delta }=0$ are

\begin{eqnarray}
\hspace{-5mm}D_{(a)} &=&\left( 4\,m_{N}^{3}-m_{N}\,{M}^{2}\right) \,\left( -{%
{m_{\pi }}}^{4}+m_{N}^{2}\,{M}^{2}\right) ,  \nonumber \\
\hspace{-5mm}D_{(b)} &=&9\,{{m_{\Delta }}}^{2}\,\left( m_{N}+{m_{\Delta }}-{M%
}\ \right) \,\left( m_{N}+{m_{\Delta }}+{M}\right) \,\left( m_{N}^{2}-{{%
m_{\Delta }}}^{2}+{{m_{\pi }}}^{2}-m_{N}\,{M}\right)  \nonumber \\
&&\times \left( -{{m_{\pi }}}^{2}+m_{N}\,{M}\right) \,\left( {{m_{\pi }}}%
^{2}+m_{N}\,{M}\right) \,\left( -{{m_{\Delta }}}^{2}+{{m_{\pi }}}%
^{2}+m_{N}\,\left( m_{N}+{M}\right) \right) ,  \nonumber \\
\hspace{-5mm}D_{(c)} &=&9\,{{m_{\Delta }}}^{2}\,\left( -4\,m_{N}^{3}+m_{N}\,{%
M}^{2}\right) \,\left( {\left( m_{N}^{2}-{{m_{\Delta }}}^{2}+{{m_{\pi }}}%
^{2}\right) }^{2}-m_{N}^{2}\,{M}^{2}\right) ,  \nonumber \\
\hspace{-5mm}D_{(d)} &=&27\,{{m_{\Delta }}}^{3}\,\left( {\left( m_{N}^{2}-{{%
m_{\Delta }}}^{2}+{{m_{\pi }}}^{2}\right) }^{2}-m_{N}^{2}\,{M}^{2}\ \right) ,
\nonumber \\
\hspace{-5mm}D_{(e)} &=&27\,{{m_{\Delta }}}^{4}\,\left( m_{N}+{m_{\Delta }}-{%
M}\right) \,\left( m_{N}+{m_{\Delta }}+{M}\right) \,\times  \nonumber \\
&&\left( {\left( m_{N}^{2}-{{m_{\Delta }}}^{2}+{{m_{\pi }}}^{2}\right) }%
^{2}-m_{N}^{2}\,{M}^{2}\right) ,  \nonumber \\
\hspace{-5mm}D_{(f)} &=&{{m_{\pi }}}^{4}-m_{N}^{2}\,{M}^{2},  \nonumber \\
\hspace{-5mm}D_{(g)} &=&9\,{{m_{\Delta }}}^{2}\,\left( {\left( m_{N}^{2}-{{%
m_{\Delta }}}^{2}+{{m_{\pi }}}^{2}\right) }^{2}-m_{N}^{2}\,{M}^{2}\right) .
\end{eqnarray}

\newpage

\bibliographystyle{npsty}
\bibliography{delta}

\end{document}